\documentclass[varenna]{cimento}

\title{Recent Developments in the Analysis of Galaxy Surveys}
\author{Will J. Percival\thanks{will.percival@uwaterloo.ca}}
\institute{Department of Physics and Astronomy, University of Waterloo, 200 University Ave W, Waterloo, ON N2L 3G1, Canada\\Perimeter Institute for Theoretical Physics, 31 Caroline St. North, Waterloo, ON N2L 2Y5, Canada}

\begin{document}

\maketitle

\begin{abstract}
  These are advanced lecture notes covering recent developments in the
  methodology used to analyse galaxy surveys. The focus is
  particularly on direct measurements of the galaxy power spectrum
  although I also discuss its Fourier transform, the correlation
  function for comparison. These 2-point statistics, under the
  assumption that the overdensity field has Gaussian statistics on
  large-scales, contain the majority of the cosmological signal
  available from the galaxy distribution. Recent developments in
  multipole measurements, dealing with systematics, convolving
  theoretical models with the survey window function, the
  approximation of covariance matrices, and weighting schemes for
  measuring evolution with redshift are considered. The focus is on
  analytic explanation of the issues involved rather than on recent
  analyses or simulation results. These notes only loosely follow the
  lectures I gave in Varenna, which were more wide ranging and
  contained more introductory material. However, I have in the
  not-too-distant-past created lecture notes for an introductory
  course on galaxy survey analysis \cite{percival13}, and I did not
  want to duplicate those notes, but instead write something new for
  these proceedings.
\end{abstract}

\tableofcontents

\section{Introduction}

This review of recent developments in the analysis of galaxy surveys
is meant to be useful for someone who already has a basic
understanding of the field. For a general introduction see, for
example, \cite{percival13}.

\section{The overdensity field}

The dimensionless overdensity field is defined as
\begin{equation}  \label{eq:delta}
  \delta(\mathbf{s}) = 
  \frac{\rho(\mathbf{s})-\langle\rho(\mathbf{s})\rangle}
    {\langle\rho(\mathbf{s})\rangle}\,,
\end{equation}
where $\rho(\mathbf{s})$ is the observed galaxy density and
$\langle\rho(\mathbf{s})\rangle$ is the expected density.

At early times, and on large-scales at present day,
$\delta(\mathbf{s})$ has a distribution that is close to that of
Gaussian, adiabatic fluctuations \cite{planck18-inflation}, and in these limits
the statistical distribution is completely described by the two-point
functions of this field.

\cite{FKP} (commonly referred to as the FKP paper) presented the first
full analysis pipeline for a galaxy survey in Fourier space. The start
of this pipeline is to define the function,
\begin{equation}  \label{eq:FKP_factor}
F(\mathbf{s})=\frac{w(\mathbf{s})}{I^{1/2}} \left[n(\mathbf{s}) - \langle
n(\mathbf{s})\rangle \right]\,,
\end{equation}
where $n(\mathbf{s})$ is the observed number density of galaxies
around location $\mathbf{s}$. The expected value
$\langle n(\mathbf{s})\rangle$ is commonly determined by means of a
synthetic catalog of {\it random} points, Poisson sampled with the
same mask and selection function as the survey. In this case,
$\langle n(\mathbf{s})\rangle \equiv \alpha n_r(\mathbf{s})$, where
$\alpha$ normalizes the weighted random catalogue with density $n_r(\mathbf{s})$
to match the weighted galaxy catalogue. The random catalogue is usually
set to have $>\sim50$ times as many points as galaxies in order that
the shot noise contribution is sub-dominant compared to that of the
galaxy catalogue.

In contrast to $\delta(\mathbf{s})$ in Eq.~(\ref{eq:delta}), the
denominator in the expression for $F(\mathbf{s})$ in
Eq.~(\ref{eq:FKP_factor}) is not a function of $\mathbf{s}$. Thus,
fluctuations in the expected density of galaxies across the survey are
not normalised out, and we have to accept that the power-spectrum of
the field $F(\mathbf{s})$ is convolved with a window function. In
general, it is not easy to divide by the density when calculating the
power spectrum as regions outside of the survey, or where no galaxies
are expected because of discreteness effects when using a random
catalogue to specific the survey selection function, would cause
divide-by-zero problems in the calculation of $F(\mathbf{s})$.

The factor $I$ normalizes this expression such that the observed monopole
moment of the power has the correct amplitude in a universe with no
window, $I\equiv \int d\mathbf{s}\,w^2{\bar{n}}^2(\mathbf{s})$.

The match between galaxy and random catalogues creates what is known
as the Integral Constraint (IC), forcing the average density within
the survey region to be zero, ignoring larger-than-survey fluctuations
in density. In Section~\ref{sec:IC-Pk}, I discuss how to calculate
$\alpha$ in such a way that the effect of this IC can be included in
models to be fitted to the data.

The FKP paper showed how the galaxies in the survey should be
optimally weighted to allow for variations in density across a survey,
balancing sample variance and shot noise. Each galaxy is weighted by
\begin{equation}  \label{eq:wght_FKP}
   w_{\rm FKP}(\mathbf{r})=\frac{1}{1+\bar{n}(\mathbf{r})\bar{P}(k)}\,,
\end{equation}
where $\bar{n}(\mathbf{r})$ is the expected density of galaxies, and
$\bar{P}(k)$ the expected power spectrum, usually fixed at a fiducial
value. This optimal weight depends on a number of assumptions -
particularly that the galaxies Poisson sample the density field, and
that the galaxies all have the same clustering strength. Revised
weighting schemes have been proposed for more realistic models of
these effects (e.g. \cite{percival04, seljak09}).

\section{Line-of-sight assumptions}

Most theoretical models (e.g. \cite{kaiser87}) make predictions for an
idealised survey in which the line-of-sight (LOS) to every galaxy is
assumed to be parallel. However, for a real galaxy survey the LOS to
different galaxies are not parallel, and analysing a survey under the
global plane-parallel assumption only gives results close to the
theory for distant surveys with small angular coverage (e.g. WiggleZ,
\cite{blake10}). For surveys covering a wide angular region (e.g. the
Baryon Oscillation Spectroscopic Survey BOSS, \cite{dawson13}), such a
global approximation gives a poor match to the standard theory.

In order to make a measurement that can be matched to theory (in which
it is assumed that LOS to galaxies are parallel), it is far better to
make a local plane-parallel approximation when analysing a survey:
that is, when measuring 2-point clustering we split the survey into
pairs of galaxies, and make a local plane-parallel approximation for
each pair. This approximation will still break down for pairs of
galaxies separated by a wide-angle, but gives results closer to the
global plane-parallel clustering amplitude for pairs whose angular
separation is small. There is a subtlety in that when analysing data,
we can choose whether to define the single line of sight as matching
that for one galaxy in a pair, or as the direction to the pair centre,
and then furthermore define how the pair centre is
calculated. However, this only produces a minor perturbation on the
overall effect.

The difference between clustering measurements made using the local
plane-parallel approximation (see next section), and the model
provided in the global plane-parallel framework, is commonly called
the wide-angle effect. This gets worse for wide-angle pairs, and the
severity of the problem depends on the distribution of pair separation
angles in a survey. Note that the plane-parallel approximation is also
commonly called the distant observer approximation.

In order to allow for wide-angle effects, one could imagine trying to
model the power spectrum calculated under the local plane-parallel
assumption using the full wide-angle theory
\cite{szapudi04,papai08}. However, it is not clear that there would be
a gain compared with using the correlation function, for which it
would be simple to split pair-counts into bins in separation,
pair-centre angle to the LOS, and angular separation, and then model
these directly.

\section{Multipole moments}

Models of the Alcock-Packynski (AP; \cite{AP}) effect and of linear
Redshift-Space Distortions (RSD; \cite{kaiser87}) show that, to first
order, under the global plane-parallel assumption, the cosmological
information of interest is contained within the first three even
power-law moments of the correlation function or power spectrum with
respect to $\mu$, the cosine of the angle that the pair or that the
Fourier mode makes with respect to the line-of-sight
(LOS). Decomposing into a Legendre polynomial basis instead of
power-law moments has the advantage of giving independent moments of
the power spectrum in the absence of a window function in the global
plane-parallel limit. Remembering that the first 3 even Legendre
polynomials are
\begin{eqnarray}
  L_0(\mu) &=& 1\,,\\
  L_2(\mu) &=& \frac{1}{2}(3\mu^2-1)\,,\\
  L_4(\mu) &=& \frac{1}{8}(35\mu^4-30\mu^2+3)\,,
\end{eqnarray}
we see that the Legendre polynomial moments can be determined by the
trivial linear combination of power-law moments. Thus we use both
interchangeably in the following depending on which is simplest to
adopt.

The Legendre polynomial moments of the unconvolved correlation
function and power spectrum in the global plane-parallel approximation
are related by
\begin{eqnarray} \label{eq:xi_l}
  \xi_\ell(s)&=& (2\ell+1) \int_0^1 d\mu\,\xi(\mathbf{s})\mathcal{L}_\ell(\mu_s)\,,\\
  P_\ell(k)&=& (2\ell+1) \int_0^1 d\mu\,P(\mathbf{k})\mathcal{L}_\ell(\mu_k)\,,
\end{eqnarray}
with inverse formulae
\begin{eqnarray}
  \xi(\mathbf{s})&=& \sum_\ell \xi_\ell(s)\mathcal{L}_\ell(\mu_s)\,,\\
  P(\mathbf{k})&=& \sum_\ell P_\ell(k)\mathcal{L}_\ell(\mu_k)\,.
\end{eqnarray}
The power spectrum and correlation function moments are related by the
Hankel transform
\begin{equation}
  P_{\ell}(k) = 4\pi i^\ell \int ds\,s^2\xi_{\ell}(s) j_{\ell}(sk)\,.
\end{equation}
In the following quantities that include the survey window function
are denoted by a prime, while unconvolved quantities are not. No
distinction is made between measured and model quantities as this
should be clear from the context.

\section{Correlation function estimators in the local
  plane-parallel formalism}  \label{sec:xi_est}

The correlation function is most commonly measured using the
Landy-Szalay estimator \cite{landy93}
\begin{equation}  \label{eq:LZ93}
  \bar{\xi}(s,\mu) = \frac{DD(s,\mu)-2DR(s,\mu)+RR(s,\mu)}{RR(s,\mu)}\,,
\end{equation}
where $\mu$ is defined with respect to the LOS to the
pair centre. $DD(s)$ is the number of galaxy-galaxy pairs within a bin
with centre $s$ normalised to the maximum possible number of
galaxy-galaxy pairs, and $RR(s)$ and $DR(s)$ are the normalised number of
random-random pairs, and galaxy-random pairs respectively. 

This estimate is biased by the integral constraint, a consequence of
the fact that the total number of galaxy-galaxy pairs is estimated
from the sample itself (equivalent to determining $\alpha$ in
Eq.~\ref{eq:FKP_factor} from the galaxies). 
\begin{equation}
  \langle 1+\bar{\xi}(s,\mu) \rangle =
  \frac{1+\xi(s,\mu)}{1+\xi_\Omega(s)}\,,
\end{equation}
where $\xi_\Omega(s)$ is the mean of the two-point correlation
function over the mask \cite{landy93}. For modern large galaxy surveys
this correction is negligibly small unless we are interested in the
clustering on very large scales. See Section~\ref{sec:IC-Pk} to see how this
affects the power spectrum measurement.

From this, we can calculate the multipoles by integrating over $\mu$
as in Eq.~(\ref{eq:xi_l}). Note that this integral should be carried out
separately from the calculation of $\xi(s,\mu)$ of Eq.~(\ref{eq:LZ93})
because the distribution of pairs in a survey is usually not uniformly
distributed in $\mu$ as required by the integral in
Eq.~(\ref{eq:xi_l}). An alternative is to define a pseudo-multipole
estimator where the kernel is not exactly the Legendre polynomial,
which generally complicates the analysis.

Note that Eq.~(\ref{eq:LZ93}) estimates the unconvolved correlation
function, and so can be directly compared with models. As we will see
later, this is not necessarily true for direct estimators of the Power
Spectrum. 

\section{Power spectrum estimators in the global plane-parallel
  formalism}

The Quadratic Maximum Likelihood (QML) estimator \cite{tegmark98}
correctly accounts for correlations between modes when optimally
measuring the power spectrum from data. In the limit of uncorrelated
modes with equal noise per mode in each bin this simplifies to the
FKP estimator \cite{FKP}.

The QML estimator is given by
\begin{equation}
  P(k_i)=\sum_{j}\mathbf{N}_{ij}^{-1}\mathbf{p}_j\,,
\end{equation}
where the power is a convolution of the inverse of a normalisation
matrix $\mathbf{N}_{ij}$ and a weighted two-point function
\begin{equation}
  \mathbf{p}_j\equiv \sum_{\alpha, \beta} F^\ast(\mathbf{k}_\alpha) \mathbf{E}_{\alpha\beta}(k_j) F(\mathbf{k}_\beta)\,.
\end{equation}
The weight is given by the estimator matrix
\begin{equation}
  \mathbf{E}(k_j)=-\frac{\partial \mathbf{C}^{-1}}{\partial P(k_j)}\,,
\end{equation}
which describes how the inverse of the density field covariance matrix
$\mathbf{C}$ changes with respect to the prior of the power spectrum
of the respective bin. If the QML normalisation is proportional to the
Fisher information, 
\begin{equation}
  \mathbf{N}_{ij} =\mathrm{tr}\left\lbrace \mathbf{C}^{-1}
    \frac{\partial\mathbf{C}}{\partial P(k_i)}\mathbf{C}^{-1}\frac{\partial\mathbf{C}}{\partial P(k_j)}\right\rbrace\,,
\end{equation}
the QML estimator is the optimal maximum likelihood estimator of the
variance of a field that obeys a multivariate Gaussian distribution
\cite{tegmark98}. I.e. assuming a Gaussian density field, the QML
estimator therefore provides an estimate of the power spectrum with
minimal errors.

Under the assumption that all modes are independent, the QML estimator
reduces to the FKP estimator
\begin{equation}
  P(k_i)=\frac{1}{N_i}\sum_{\mathbf{k}_\alpha\in\mathrm{bin}\,i}\left\vert F(\mathbf{k}_\alpha)\right\vert^2\,,
\end{equation}
which simply averages the power in the Fourier modes. The FKP-style
estimator is commonly applied even when the assumptions required for
optimality are not valid. This concept of averaging rather than
performing the optimised combination of on and off-diagonal modes is
also used in the estimator in the local plane-parallel formalism given
in the next section.

\section{Power spectrum estimators in the local plane-parallel
  formalism} \label{sec:fourier}

Following the ethos behind the FKP power spectrum estimator, in the
local plane-parallel approximation, we can define as the statistic
that we want to reduce the data to as the order-$n$ power-law moments
of the window-convolved power spectrum \cite{yamamoto06}
\begin{equation}  \label{eq:Pn-start}
  P'_n(k) = \frac{1}{4\pi} \int d\Omega_k \int d^3s_1\,\int d^3s_2\,
    (\hat{\mathbf{k}}\cdot\hat{\mathbf{s_1}})^{n} 
    F(\mathbf{s_1}) F(\mathbf{s_2}) 
    e^{i\mathbf{k}\cdot(\mathbf{s}_1-\mathbf{s}_2)}-P_s\,,
\end{equation}
where we have assumed that the LOS of the pair of galaxies lies along
direction $\mathbf{s}_1$, $d\Omega_k$ is the solid angle element in
$k$-space, and we denote the window convolved power spectrum
$P'$. $P_s$ is the shot noise term. The local plane-parallel
approximation is used to both define a single LOS to each pair of
galaxies, and to define that LOS as the direction to one of the
galaxies.

By writing the LOS in terms of only $\mathbf{s}_1$, we can split the
integrals in Eq.~(\ref{eq:Pn-start}), such that 
\begin{equation}  \label{eq:Pn-An}
  P'_n(k)=\frac{1}{4\pi} \int d\Omega_k A'_n(\mathbf{k})[A'_0(\mathbf{k})]^*-P_s\,,
\end{equation}
where
\begin{equation}  \label{eq:An}
  A'_n(\mathbf{k}) = \int d^3s\,F(\mathbf{s}) (\hat{\mathbf{k}}\cdot\hat{\mathbf{s}})^n
    e^{i\mathbf{k}\cdot\mathbf{s}}\,.
\end{equation}
\cite{bianchi15} and \cite{scoccimarro15} showed that these can be
solved using Fast Fourier Transforms (FFTs) on a Cartesian grid after
substituting the trivial decomposition
\begin{equation}  \label{eq:kdotr}
  \hat\mathbf{k}\cdot\hat\mathbf{s}=\frac{k_x s_x+k_y s_y+ k_z s_z}{ks}\,,
\end{equation}
into Eq.~(\ref{eq:An}).  Recently, \cite{hand17} proposed a Legendre
polynomial decomposition of $\hat\mathbf{k}\cdot\hat\mathbf{s}$ which
allowed fewer FFTs to be used than in the simple approach above. Many
FFT libraries are available, with FFTW being a commonly used example
({\tt http://www.fftw.org/}). Thus this statistic can be quickly
measured for a given catalogue.

\section{Grid assignment, aliasing and interlacing}

When using FFTs to Fourier transform $F(\mathbf{s})$
(Eq.~\ref{eq:FKP_factor}) we must sample $F(\mathbf{s})$ on a
grid. Small-scale modes, unresolved due to the finite grid can alias
large-scale modes, leading to the wrong power spectrum
measurement. One way to avoid this is to use direct summation rather
then FFT for the Fourier transform. However this is slow for large
galaxy surveys, particularly when applied to a large random catalogue.

The effect of aliasing can be reduced through the choice of grid
assignment scheme to interpolate $F(\mathbf{s})$ onto the regular
grid, and by use of interlacing. \cite{sefusatti16} provided an
excellent review of these issues, and is the source for the ideas
presented in this section. Considering calculating $F(\mathbf{s})$ at
the grid points $\mathbf{s}_j$, we see that we can assign galaxies to
$n(\mathbf{s})$ as given by
\begin{equation}
  n(\mathbf{s}_j) = \frac{1}{H^3}\sum_{i=1}^{N_{\rm gal}}\, W^{(p)}(\Delta s_x/H)\,
  W^{(p)}(\Delta s_y/H)\,W^{(p)}(\Delta s_z/H)\,,
\end{equation}
where $\Delta\mathbf{s}=\mathbf{s}_j-\mathbf{s}_i=(\Delta s_x,\Delta s_y,\Delta s_z)$, and
$H$ is the grid size. If a random catalogue is used to define the
survey mask, then the assignment of these points to the grid follows
the same procedure. 

It is common to consider piecewise polynomial functions for the
1-dimensional functions $W^{(p)}$, which simply correspond to convolving
a top-hat function with itself $(p-1)$ times:

\noindent
Nearest Grid Point (NGP)
\begin{equation}
  W^{(1)}(t)=\left\{
    \begin{array}{ll}
    1 & \mathrm{for}~ |t|<\frac{1}{2} \\
    0 & \mathrm{otherwise}
    \end{array}
  \right.
\end{equation}
\noindent
Cloud-In-Cell (CIC)
\begin{equation}
  W^{(2)}(t)=\left\{
    \begin{array}{ll}
    1-|t| & \mathrm{for}~ |t|<1 \\
    0 & \mathrm{otherwise}
    \end{array}
  \right.
\end{equation}
\noindent
Triangular Shaped Cloud (TSC)
\begin{equation}
  W^{(3)}(t)=\left\{
    \begin{array}{ll}
    \frac{3}{4}-t^2 & \mathrm{for}~ |t|<\frac{1}{2} \\
    \frac{1}{2}\,\left(\frac{3}{2}-|t|\right)^2 & \mathrm{for}~ \frac{1}{2}\le |t|<\frac{3}{2} \\
    0 & \mathrm{otherwise}
    \end{array}
  \right.
\end{equation}
\noindent
Piecewise Cubic Spline (PCS)
\begin{equation}
  W^{(4)}(t)=\left\{
    \begin{array}{ll}
    \frac{1}{6}\,\left(4-6\,t^2+3\,|t|^3\right) & \mathrm{for}~ 0\le |t|< 1 \\
    \frac{1}{6}\,\left(2-|t|\right)^3 & \mathrm{for}~ 1 \le |t|< 2 \\
    0 & \mathrm{otherwise}
    \end{array}
  \right.
\end{equation}

The interpolation function acts as a convolution in
configuration-space, and hence is a multiplicative factor in Fourier
space and can be removed by dividing $F(\mathbf{k})$ by the Fourier
transform of the window, $W^{(p)}(k_x)\,W^{(p)}(k_y)\,W^{(p)}(k_z)$,
where
\begin{equation}
  W^{(p)}(k)=\left[\frac{\sin(kH/2)}{(kH/2)}\right]^p\,.
\end{equation}
While this corrects for the magnitude of the convolution its effects
live-on in the amplitude of the aliasing effect.

A method to partially correct for aliasing based on the interlacing of
two grids is discussed in the classic text of \cite{hockney81}. The
key idea is to perform an additional, configuration-space
interpolation onto a grid shifted by $H/2$ in all spatial directions,
and then take the average of the two
\begin{equation}
  F(\mathbf{k})=\frac{1}{2}\left[F_1(\mathbf{k})+F_2(\mathbf{k})\right]\,,
\end{equation}
where $F_1(\mathbf{k})$ and $F_2(\mathbf{k})$ represent the individual
transforms. This removes the leading order aliasing terms. 

\section{Linking Fourier and Fourier-Bessel bases}

The link between a Fourier-space decomposition and a decomposition
into a basis consisting of spherical harmonics and spherical Bessel
functions (hereafter known as a Fourier-Bessel basis) is given by the
Rayleigh expansion of a plane wave. In terms of Legendre polynomials
this is written:
\begin{equation}
  e^{i\mathbf{k}\cdot\mathbf{s}} = \sum_\ell i^\ell(2\ell+1)j_\ell(ks)
    \mathcal{L}_\ell(\hat{\mathbf{k}}\cdot\hat{\mathbf{s}}),
\end{equation}
and in terms of Spherical Harmonics:
\begin{equation}  \label{eq:planewave}
  e^{i\mathbf{k}\cdot\mathbf{s}} = 4\pi \sum_{\ell m} i^\ell
    j_\ell(ks) Y_{\ell m}(\hat{\mathbf{k}}) Y^*_{\ell m}(\hat{\mathbf{s}}).
\end{equation}

Using this, we can see how Eq.~(\ref{eq:An}) can be solved using a
Fourier-Bessel basis. As explained by \cite{datta07}, consider
differentiating Eq.~(\ref{eq:planewave}) with respect to $ks$ $n$
times to give:
\begin{equation}  \label{eq:planewave_deriv}
  i^n (\hat{\mathbf{k}}\cdot\hat{\mathbf{s}})^n e^{iks(\hat{\mathbf{k}}\cdot\hat{\mathbf{s}})} 
  = 4\pi \sum_{\ell m} i^\ell
  j^{(n)}_\ell(ks) Y_{\ell m}(\hat{\mathbf{k}}) Y^*_{\ell m}(\hat{\mathbf{s}}),
\end{equation}
where $j_\ell^{(n)}(ks)$ is the $n$'th derivative of the spherical Bessel
function with respect to $ks$. 

\cite{datta07} used the trick of taking the derivative of the plane
wave expansion to directly link power spectrum multipoles to a
Fourier-Bessel decomposition. This was applied to the linear model RSD
to write $\delta_{\ell m}(k)$ in terms of expansions in spherical
Bessel functions and their derivatives. The work of \cite{datta07}
was applied by \cite{castorina18a,castorina18b} to investigate the
impact of wide-angle effects from Fourier-based multipole
measurements.

To see how this trick can allow us to measure the statistic in
Eq.~(\ref{eq:An}), using a Fourier-Bessel rather than Fourier basis,
we substitute the plane-wave expansion into Eq.~(\ref{eq:An}) to give
\begin{equation}
  A'_n(\mathbf{k}) = 4\pi \int d^3s F(\mathbf{s}) \sum_{\ell m} i^{\ell-n}
      j^{(n)}_\ell(ks) Y_{\ell m}(\hat{\mathbf{k}}) Y^*_{\ell m}(\hat{\mathbf{s}})\,.
\end{equation}
If we also expand $F(\mathbf{s})$ in a basis of Spherical Harmonics and the
$n$'th derivative of spherical Bessel functions
\begin{equation}  \label{eq:newexpansion}
  F^{(n)}_{\ell m}(k) = \int \mathrm{d}^3s\, F(\mathbf{s})
    Y_{\ell m}^\star(\hat{\mathbf{s}}) j^{(n)}_\ell(ks)\,,
\end{equation}
then we can write $A_n(\mathbf{k})$ in terms of $F^{(n)}_{\ell m}(k)$ as
\begin{equation}  \label{eq:An-Fourier-Bessel}
  A'_n(\mathbf{k}) = 4\pi \sum_{\ell m} i^{\ell-n} 
    F^{(n)}_{\ell m}(k) Y_{\ell m}(\hat{\mathbf{k}})\,.
\end{equation}

To intuitively see how the Fourier and Fourier-Bessel solutions for
$A_n(\mathbf{k})$ are related, note that derivatives of spherical
Bessel functions can be rewritten as the difference between standard
spherical Bessel functions. For example:
\begin{equation}  \label{bessel-derivative}
  j^{(1)}_\ell(ks)=-j_{\ell+1}(ks)+\frac{\ell}{ks}⁢j_\ell⁡(ks).
\end{equation}
Subsequent derivatives can be related to standard spherical Bessel
functions by recursive application of this formula. As $\ell/(ks)$ is
equal to $k_\perp/k$ defined locally at position $\mathbf{s}$ for
modes of wavenumber $\ell$ in a Fourier-Bessel function decomposition,
we can see that taking the derivatives in
Eq.~(\ref{eq:planewave_deriv}) is directly related to the local
multipole expansion as considered in
Eq.~(\ref{eq:kdotr}). I.e. Instead of using the derivatives of the
spherical Bessel functions to get Eq.~(\ref{eq:An-Fourier-Bessel}), it
would have instead been possible to use the local definition of
$\hat{\mathbf{k}}\cdot\hat{\mathbf{s}}$ in the integral to get the
same result.

Substituting Eq.~(\ref{eq:An-Fourier-Bessel}) for $n$ and $n=0$ into
Eq.~(\ref{eq:Pn-An}), and using the orthogonality relations for
spherical harmonics removes the angular integral and gives the simple
result that
\begin{equation}
  P'_n(k) = 4\pi\sum_{\ell m}F^{(n)}_{\ell m}(k) [F^{(0)}_{\ell m}(k)]^*-P_s\,.
\end{equation}

Thus we see that we can use either Fourier or Fourier-Bessel bases to
measure $P'_n(k)$ from a galaxy redshift survey under the local
plane-parallel approximation. The use of a Fourier-Bessel basis does
not alleviate wide-angle effects: these are built in to the power
spectrum multipole definition (Eq.~\ref{eq:Pn-start}). Indeed,
wide-angle effects are fundamentally model dependent, so we cannot
remove them completely using a model-independent or fiducial-model
based estimator.

Given the same end point for the measurements, the Fourier approach is
preferred as it allows the use of FFTs to perform the transforms
required, saving computational resources. Using a Fourier-Bessel basis
may help when calculating the cross-power spectrum of a galaxy
redshift survey with an angular survey.

\section{Window convolution of models}

Given that we measure the power spectrum convolved with the window
function, we need a fast mechanism to convolve models to be compared
to $P'_n(k)$. \cite{wilson17} showed that we can perform the required
3D convolution quickly using the Hankel transform relation between the
window-convolved multipole moments in configuration and Fourier space,
\begin{equation}  \label{eq:hankel-conv}
  P'_{\ell}(k) = 4\pi i^\ell \int ds\,s^2\xi'_{\ell}(s) j_{\ell}(sk)\,.
\end{equation}
Crucially, this equation holds for both the unconvolved and convolved
power spectrum and correlation function pairs in both the global
plane-parallel \cite{wilson17} and local plane-parallel
\cite{beutler17} limits. The Hankel tranform can be quickly solved
using a 1D FFT, although care has to be taken given the oscillatory
nature of the integrand, as described in \cite{hamilton2000}.

By using this transform we can calculate the convolved model power
spectra using multiplications in real space. The Legendre moments of
the convolved correlation function are defined
\begin{equation}  \label{eq:xip1}
  \xi'_\ell(s) = \frac{2\ell + 1}{4\pi}\int d\Omega_s\,
    \xi(\mathbf{s}) W^2(\mathbf{s})\mathcal{L}_{\ell}(\hat{\mathbf{s}} \cdot \hat{\mathbf{x}}_1)\,,
\end{equation}
where $\xi(\mathbf{s})$ and $W^2(\mathbf{s})$ are the anisotropic
correlation function and window function. $\mathcal{L}_{\ell}$ is the
Legendre polynomial of order $\ell$, here written as a function of the
LOS to one galaxy in each pair $\mathbf{x}_1$, matching the
assumptions of Eq.~(\ref{eq:Pn-start}).

We define the moments of the window function as
\begin{equation}
  W_p^2(s)=\frac{2p+1}{4\pi} \int d\Omega_s \int d\mathbf{x}_1 
    W(\mathbf{x}_1)W(\mathbf{x}_1+\mathbf{s})\mathcal{L}_p(\mu_s)\,,
\end{equation}
which can be calculated by using the random catalogue to perform the
integrations with a Monte-Carlo based technique. Substituting this
into Eq.~(\ref{eq:xip1}), and also expanding the unconvolved
correlation function in Legendre moments means that we can rewrite
this equation as
\begin{equation}  \label{eq:xip2}
  \xi'_{\ell}(s) = (2\ell + 1)\sum_{L}\xi_{L}(s) \sum_p\frac{1}{2p + 1}W_p^2(s) a^{p}_{L\ell}\,,
\end{equation}
where $a^{p}_{L\ell}$ are the solutions to the Equation
$\mathcal{L}_{\ell}(\mu)\mathcal{L}_{L}(\mu) = \sum_p a^{p}_{\ell
  L}\mathcal{L}_p(\mu)$, and can be obtained by substituting in the
polynomials and equating powers of $\mu$. The window convolution
spreads the linear information to $\xi'_{\ell}(s)$ with $\ell>4$ and
the $\xi'_{\ell}(s)$ modes with $\ell\le4$ depend on higher order
moments. However, it is common to only fit to the first three even
multipoles, ignoring the potential information at higher orders, and
furthermore only apply the window convolution to the linear model,
which is reasonable as the window effect diminishes to smaller
scales. Expanding Eq.~(\ref{eq:xip2}) gives that the relevant
expansion components are 
\begin{eqnarray}
  \xi'_0(s) &=& \xi_{0}W_{0}^2  + \frac{1}{5}\xi_2W^2_2 + \frac{1}{9}\xi_4W^2_4\,,\\
  \xi'_2(s) &=& \xi_{0} W_{2}^2 + \xi_2\left[W^2_0 +
      \frac{2}{7}W^2_2 + \frac{2}{7}W^2_4\right]
      +\xi_4\left[\frac{2}{7}W^2_2 + \frac{100}{693}W^2_4 +
      \frac{25}{143}W^2_6\right]\,,\\
  \xi'_4(s) &=& \xi_{0} W_{4}^2 + \xi_2\left[\frac{18}{35}W^2_2
       +\frac{20}{77}W^2_4 +\frac{45}{143}W^2_6\right]\nonumber\\
  &&+\xi_4\bigg[W^2_0 + \frac{20}{77}W^2_2 + \frac{162}{1001}W^2_4 + 
       \frac{20}{143}W^2_6 + \frac{490}{2431}W^2_8 \bigg]\,,
\end{eqnarray}
keeping $\xi_\ell$ terms with $\ell\le4$, and including all of the
relevant window multipole moments. This matches the premise that
linear theory is complete to $\ell=4$, but the window function has no
such constraint. Given $\xi'_\ell(s)$, the model power spectrum multipoles
can be calculated using Eq.~(\ref{eq:hankel-conv}).

\section{Power spectrum Integral Constraint} \label{sec:IC-Pk}

This formalism for the window also makes it easy to see how the
integral constraint can be included in models \cite{beutler17}. For
the power spectrum, the integral constraint is relevant because, to
formulate $F(\mathbf{s})$ as in Eq.~(\ref{eq:FKP_factor}), we have
matched $\langle n(\mathbf{s})\rangle$ to the actual observed density
of galaxies. We can assume that the variations in the expected
distribution of $\langle n(\mathbf{s})\rangle$ as a function of
$\mathbf{s}$ are known, but that the normalisation is incorrect, so
\begin{equation}
  \langle n(\mathbf{s})\rangle_\mathrm{assumed} 
    = (1+C)\langle n(\mathbf{s})\rangle_\mathrm{true}\,,
\end{equation}
where $C$ is a constant. The multiplicative nature of $C$ with
$\langle n(\mathbf{s})\rangle$ means that the constant is inside the
window function convolution and is equivalent to an additive
contribution to $\xi_0$. We ignore the possibility that a mistake has
also been made in the calculation of $I$.

The size of the correction depends on how the randoms have been
matched to the galaxies. If the total number of weighted pairs has
been matched, then we have forced that $P'_0(0)=0$. In this case, the model to
be compared to the data is 
\begin{equation}
  P'_{\ell,\mathrm{ic-corrected}}(k) = P'_\ell(k)-\frac{P_0'(k=0)}{W_0^2(k=0)}W_\ell^2(k)\,,
\end{equation}
where
\begin{equation}
  W^2_{\ell}(k) = 4\pi\int ds\,s^2W_{\ell}^2(s)j_{\ell}(sk)\,.
\end{equation}
Considering this from a different stand-point, by defining $\alpha$ by
matching the total number of weighted pairs in the estimator, we have
a simple expression for the integral constraint to be included in the
model to match the measurement. Thus this is the preferred method for
calculating $\alpha$.

\section{Covariance matrix under Gaussian assumption}

In order to make statistical inferences from the measured power, we
need to model the distribution from which it is drawn. It is common to
assume that the power spectrum multipoles are drawn from a
multi-variate Gaussian population, in which case the Likelihood for
the power spectrum is
\begin{equation}  \label{eq:like1}
  \mathcal{L}(\mathbf{x}|\mathbf{p},\Psi_{\mathrm{true}})
    =\frac{|\Psi_{\mathrm{true}}|}{\sqrt{2\pi}}\exp
    \left[-\frac{1}{2}\chi^2(\mathbf{x},\mathbf{p},\Psi_{\mathrm{true}})\right]\,,
\end{equation}
where
\begin{equation}
  \chi^2(\mathbf{x},\mathbf{p},\Psi_{\mathrm{true}}) \equiv 
    \left[\mathbf{x}^d-\mathbf{x}(\mathbf{p})\right]\Psi_{\mathrm{true}}
    \left[\mathbf{x}^d-\mathbf{x}(\mathbf{p})\right]\,.
\end{equation} 
In the example considered in these notes, the data $\mathbf{x}^d$, and
model for the data $\mathbf{x}(\mathbf{p})$, would be the power
spectra (or correlation function), with the parameter $\mathbf{p}$
being the cosmological parameters of interest. $\Psi_{\mathrm{true}}$
is the true inverse covariance matrix.

For power spectrum measurements the covariance is
\begin{equation}
  \mathrm{Cov}\left[P'_n(k_i), P'_{n'}(k_j)\right]
    = \langle P'_n(k_i)\,P'_{n'}(k_j)\rangle - \langle P'_n(k_i) \rangle \langle P'_{n'}(k_j) \rangle\,,
\end{equation}
where $P'_n(k_i)$ is the window-convolved power-law moment of the
power spectrum, calculated in the local plane-parallel approximation,
and binned into $k$-bin $i$. Each of the window convolution (including
weighting), local plane-parallel geometry, power-law moment and
binning effects will complicate the covariance from the simple form
without these, which is given by
\begin{equation}
  \mathrm{Cov}\left[P(\mathbf{k}),\,P(\mathbf{k}')\right]
    = \frac{2}{V}\left[P(\mathbf{k})+\frac{1}{\bar{n}}\right]^2
    \delta^D(\mathbf{k}-\mathbf{k}')\,.
\end{equation}
Here, $\delta^D $ is the Dirac delta function and the shot noise term
assumes that the galaxies Poisson sample the underlying matter
field. $V$ is the volume of the survey. This expression follows from
Wick's theorem for a Gaussian random field with zero mean
\begin{equation}
  \langle \delta_1\,\delta_2\,\delta_3\,\delta_4 \rangle = 
  \langle \delta_1\,\delta_2 \rangle \langle \delta_3\,\delta_4 \rangle +
  \langle \delta_1\,\delta_3 \rangle \langle \delta_2\,\delta_4 \rangle + 
  \langle \delta_1\,\delta_4 \rangle \langle \delta_2\,\delta_3 \rangle\,,
\end{equation}
and the standard expression for the variance of a Poisson sampling.

For the local plane-parallel power spectrum estimator of
Eq.~\ref{eq:Pn-start}, applying Wick's theorem to $A'_n(\mathbf{k})$
leads to
\begin{eqnarray}
  \mathrm{Cov}\left[P'_n(k_i), P'_{n'}(k_j)\right]
    &=& \frac{1}{(4\pi)^2} \int_i d\Omega_k \int_j d\Omega_{k'} 
        \langle A'_n(\mathbf{k})A'_{n'}(\mathbf{k'})^*\rangle 
        \langle A'_0(\mathbf{k})A'_0(\mathbf{k'})^*\rangle\\\nonumber
    &+& 
        \langle A'_n(\mathbf{k})A'_0(\mathbf{k'})^*\rangle 
        \langle A'_0(\mathbf{k})A'_{n'}(\mathbf{k'})^*\rangle\,.
\end{eqnarray}
No way of writing this equation in a form that allows its calculation
using FFTs and Hankel transforms has yet been found without also
applying simplifying assumptions. A common assumption to make is that
the power spectrum is constant over the extent of the window function,
so that the convolution breaks-down
\cite{FKP,howlett17,blake18}. \cite{blake18} showed that this results
in a form for the covariance that can be solved using only FFTs.

Because of these complications it is common to estimate the covariance
matrix in a brute-force way, from a large set of mock catalogues that
match the survey geometry and analysis method
\begin{equation} \label{eq:cov-from-mocks}
  \mathrm{Cov}\left[P'_n(k_i), P'_{n'}(k_j)\right] 
  = \frac{1}{N_m - 1} \sum^{N_m}_{m=1}
  \left[P'_{n,m}(k_i)-\overline{P'}_{n}(k_i)\right]
  \left[P'_{n',m}(k_j)-\overline{P'}_{n'}(k_j)\right]\,,
\end{equation}
where the sum is over $N_m$ mock catalogues, and $\overline{P'}$ is
the mean power spectrum over those mocks. This method automatically
includes all of the linear effects discussed above, and can also
include non-linear effects to the extent that they are included in the
method used to create the mocks, and the Gaussian assumption
holds. Given fast methods for creating the mocks and complications due
to non-linear shot-noise, galaxy bias and Redshift Space Distortions
(RSD) there are many advantages to such an approach.

The disadvantages include that this covariance matrix lacks
fluctuations caused by $k$-modes larger than the simulation box -
so-called supersample covariance \cite{takada13}. This can be included
by adjusting the mocks to force each to have slightly different
cosmological parameters \cite{howlett17}.

An additional problem is that the estimate of the covariance matrix
has errors that systematically distort the likelihood. Formally,
Eq.~(\ref{eq:cov-from-mocks}) gives an estimate drawn from a Wishart
distribution. We can allow for the bias induced using a perturbative
analysis \cite{dodelson13,taylor13}, which has to be adjusted if
parameter errors are estimated from the likelihood surface
\cite{percival14}. Alternatively, as demonstrated by
\cite{sellentin16}, one should perform a joint likelihood analysis of
both the power spectrum and covariance matrix. With the assumption of
a Jeffreys prior allowing us to use Bayes theorem to determine the
distribution of the true covariance matrix given our estimate from
mocks, we need to adjust the Likelihood of Eq.~(\ref{eq:like1}), to
\begin{equation}
   \mathcal{L}(\mathbf{x}^d|\mathbf{x}(\mathbf{p}),\Psi) \propto
   \left[1+\frac{\left[\mathbf{x}^d-\mathbf{x}(\mathbf{p})\right]\Psi
       \left[\mathbf{x}^d-\mathbf{x}(\mathbf{p})\right]}{N_m-1}
  \right]^{-\frac{N_m}{2}}\,.
\end{equation}
This offers a neater and statistically more rigorous method for
correcting for the approximate covariance matrix of
Eq.~(\ref{eq:cov-from-mocks}) compared with the perturbative solution.

\section{1-point Systematics}  \label{sec:1-point-sys}

The measurement of the power spectrum from a galaxy survey is likely
to be contaminated by systematic effects that alter the observed
galaxy density such that the fluctuations are not driven only by
astrophysical processes. Due to the way most galaxy redshift surveys
to date have been constructed by spectroscopic follow-up observation
of targets selected from imaging data, it is natural to think that
there is a split in contaminants between angular and radial
directions. However, this is not necessarily the case: for example,
removing faint targets in a patch of the sky would tend to remove
high-redshift galaxies in an apparent magnitude-limited
sample. Consequently, we do not make any such separation here.

Using the observed target distribution, coupled with maps of the
distribution of causes of potential problems - for example, imaging
depth maps, or maps of bright star locations - one can look for
fluctuations in target density. We expect the cosmological
fluctuations to be independent of the systematics, and so any
statistically significant correlation is indicative of a problem in
the sample. \cite{ross11,ross12} undertook a careful analysis of
potential systematics in BOSS, and developed a set of multiplicative
weights that correct the galaxy density for these fluctuations, in
effect creating a different window for the galaxies compared to the
random catalogue, so that Eq.~(\ref{eq:FKP_factor}) is changed to
\begin{equation}  
F(\mathbf{s})=\frac{w(\mathbf{s})}{I^{1/2}} [w_{\mathrm sys}n(\mathbf{s}) - \langle
n(\mathbf{s})\rangle ]\,,
\end{equation}
where $w_{\mathrm sys}$ are the systematic weights. These weights
increase the noise in our estimator. This would be reduced by
weighting instead the randoms to match the galaxies
\cite{bautista18}. To see this, consider the toy example of a Poisson
distribution with mean and variance $N$, weighted by a set of weights
with variance $\sigma_w^2$. The variance of the weighted sample is
$N(1+\sigma_w^2)$, and so is always greater than the unweighted field.

Weighting the randoms rather than the galaxies also shows that this
multiplicative weighting is equivalent to an additive contaminant
\cite{kalus18}
\begin{equation}  
  F(\mathbf{s})=\frac{w(\mathbf{s})}{I^{1/2}} [n(\mathbf{s}) - \langle
    n(\mathbf{s})\rangle ]+f_{\mathrm sys}(\mathbf{s})\,.
\end{equation}
The benefit of writing the effect like this is that we can equate the
weighting applied to correct for systematics to the mode deprojection
technique. Mode deprojection works by setting the covariance of modes
to be removed to be infinite in the covariance matrix of the unbinned
power
\begin{equation}
  \mathrm{Cov}\left[P'(\mathbf{k}), P'(\mathbf{k}')\right]
  \rightarrow
  \mathrm{Cov}\left[P'(\mathbf{k}), P'(\mathbf{k}')\right]
  +\lim_{\sigma\rightarrow\infty}\sigma f_{\mathrm sys}(\mathbf{k})f_{\mathrm sys}(\mathbf{k}')^*\,.
\end{equation}
\cite{kalus16} showed that this is mathematically equivalent to
weighting the randoms, modulo making the correct normalisation when
calculating the power spectrum. The standard procedure can easily
be modified to include the required renormalisation when measuring the
power - adjusting the effective number of modes as required.

It is possible to extend these ideas to remove multiple contaminants,
and to remove sets of contaminants that span a space where systematic
errors are suspected. However, as discussed in \cite{kalus18}, the
problem in general lies not in removing the contaminants, but knowing
which modes are affected: the removal of contaminants only works for
known unknowns, and fails for unknown unknowns.

\section{2-point Systematics}

We now consider a common problem induced by the mechanics of fiber-fed
multi-object spectrographs. Most have a physical limit on how close
the ends of fibers can be placed in the focal plane of the telescope
such that they cannot observe close targets in a single pass of the
instrument on the sky. Thus there is a geometrical difference between
the samples selected for observation and the parent sample from which
it was selected. Furthermore, in a sample of galaxies, close pairs
tend to have a higher bias as they are located in higher mass haloes
compared with isolated galaxies. Thus the lack of close pairs of
galaxies due to fibre collisions changes the clustering between
observed and parent samples even at large separations due to the
change in mean bias. This would not be a problem if this difference
occurred uniformly across the sky as we would then be simply selecting
a lower bias galaxy population compared with the full target sample. However,
this lack of pairs is often avoided in regions of overlapping
observations, leading to an anisotropic mean bias in the observed
sample. In addition, obviously, the small-scale clustering is strongly
affected as we lose small-separation pairs, such that there are no
angular pairs with separation smaller than the instrumental cut-off in
1-pass regions.

The problem described above is inherently of higher order than the
issues raised in Section~\ref{sec:1-point-sys}. In that section, we
considered issues that were equivalent to changing the window through
which the survey was observed. In contrast, close-pair effects are
2-point in origin as they depend on the overdensity at two
positions: we cannot observe one galaxies if there is another
nearby. To correct for these, we cannot easily use the techniques
described in Section~\ref{sec:1-point-sys}. Surveys such as DESI
\cite{desi1,desi2} have more complicated, but related problems due to
experimental limitations on how the fibres can be placed in the focal
plane of the telescope.

\cite{bianchi17} proposed the Pairwise Inverse Probability (PIP)
method to correct for the kind of 2-point systematic arising from
hardware limitations. For this we need a complete parent sample from
which the observed subsample is selected. The PIP method estimates the
probability that a pair of objects can be observed by counting how
many times it is observed in a set of possible surveys selected for
observation from the parent sample. This set can be created by
translating or rotating the survey, or by rerunning the algorithm used
to select the observed subsample with different randomly chosen
priorities for different objects. The key thing is that all surveys in
the set are equally likely, and the number of pairs of objects in the
parent sample that are never observed in any realisation of the survey
is negligible.

By weighting each pair by the inverse of this probability when
counting pairs in order to estimate the correlation function, we
recover pair counts with the same expected value as those of the full
parent sample. As an example, consider the situation where close pairs
are only observed in parts of the survey. In order to ensure that
there are no zero-probability pairs in the parent catalogue, we need
to move the survey when determining the set of samples, so that any
close-pair in the parent has a chance of falling into an overlap
region in some surveys in the set. Because they are only observed in
selected regions, close-pairs will be given a lower probability than
wide-separation pairs, leading the counts to be upweighted in the sums
for any realisation, correcting for this effect.

One issue with the technique is the time it takes to perform the
calculation. For a galaxy survey with $10^7$ galaxies, there are
$10^{14}$ pairs, and if we create a set of $10^3$ possible survey
realisations, the PIP calculation is of order $10^{17}$. The
computational burden can be minimised by calculating the weights
on-the-fly while pair counting to estimate the correlation function,
based on storing the selection of galaxies in each survey in a
bit-wise way (it's a yes/no decision on whether each object in the
parent makes it into a particular survey). The weight can then be
quickly calculated using a bitwise sum \cite{bianchi17}.

\cite{percival17} showed how the angular clustering measurement in the
parent can be incorporated into the method to improve signal, while
\cite{bianchi18} and \cite{mohammad18} showed that the method works
for the Dark Energy Spectroscopic Survey mocks and the VIMOS Public
Extragalactic Redshift Survey (VIPERS), respectively.

A similar method to debias power spectrum measurements using only FFTs
and Hankel transforms has yet to be developed, and it would also be
useful to have a method to correct the overdensity field as used in
reconstruction \cite{eisenstein07} for such effects.

\section{Binning in redshift and redshift-dependent weighting}

Future surveys such as DESI \cite{desi1,desi2} and Euclid
\cite{laureijs11} will cover a wide range in redshift, such that there
will be significant evolution in the populations of galaxies
observed. Thus we either need to allow for this evolution when
analysing the data, or divide the survey in redshift prior to
analysis. Dividing galaxies based on their redshifts into shells will
tend to miss pairs of galaxies where galaxies are in different
bins. It would also be possible to split instead by radial
pair-centre rather than galaxy position, which also mitigates for the
effect of the window on RSD measurements \cite{nock10}.

An alternative is to perform multiple analyses of the full sample
using sets of weights optimised to measure the evolving quantities of
interest. I.e. binning can be seen as using a set of top-hat weights
in redshift, and this is not necessarily the optimal choice. Using
Fisher matrix based techniques, one can find sets of weights optimised
for BAO \cite{zhu15}, RSD \cite{ruggeri17a}, and primordial
non-Gaussianity \cite{mueller17} measurements. 

These ideas were recently applied to the quasar sample in the
extended-Baryon Oscillation Spectroscopic Survey (eBOSS), recovering
BAO and RSD based measurements on evolving parameters using multiple
analyses with different sets of weights
\cite{ruggeri17b,zhu18,wang18,ruggeri18}.

\section{Reconstruction}

While the bulk motion of material in the Universe drives structure
growth, it also acts to smooth the primordial overdensity field,
leading to the degradation of the BAO feature on small scales. The
basic idea behind reconstruction is to move the late-time
over-densities back to their initial positions, sharpening the
BAO peak \cite{eisenstein07}. In terms of information, the bulk motion
moves the small-scale 2-point information into higher order terms, and
reconstruction recovers this information \cite{schmittfull15}. 

Reconstruction requires us to know the displacement field linking
Eulerian and Lagrangian positions. This displacement field can be
approximated using the standard Zel’dovich displacements,
\begin{equation}  \label{eq:zeld-recon}
  \nabla \cdot \mathbf{\Psi} + \frac{f}{b} 
    \nabla\left(\mathbf{\Psi\cdot \hat{r}}\right)\mathbf{\hat{r}} = -\frac{\delta_g}{b}\,.
\end{equation}
where $\mathbf{\Psi}$ is the displacement field, $f=d\ln D(a)/d\ln a$,
$D(a)$ is the linear growth rate, and $\sigma_8$ normalises the
amplitude of the linear power spectrum.

Solving Eq.~(\ref{eq:zeld-recon}) is complicated by the LOS-dependent
RSD term, and that the RSD field has a non-zero curl component. Two
approaches have been proposed: solving this equation on a grid
spanning the survey using finite difference techniques
\cite{padmanabhan12}, and a FFT based technique that iteratively
solves for the LOS-dependence \cite{burden15}. Both of these
techniques have been successfully applied to the analysis of data
(e.g. \cite{alam17}).  Eq.~\ref{eq:zeld-recon} is solved after
smoothing the observed field in order to focus on the large-scale bulk
motions. This changes the shape of the recovered power spectrum,
requiring changes in BAO fitting routines \cite{seo16}.

It is easy to imaging that we can do better than solving
Eq.~\ref{eq:zeld-recon} as a way to recover the bulk motions. For
example there is extra information available - such as that the
initial distribution of over-densities was homogeneous. The
development of algorithms to reconstruct the initial density-field
from an evolved field has a long history (e.g. \cite{narayanan99}),
stretching back even before the improvement of BAO observations was
considered. Many methods have been proposed. Recent highlights
include: Iterative reconstruction \cite{schmittfull17,hada18}, which
removes the need to specify a smoothing scale. More complicated
schemes have been proposed based on limiting the information used,
allowing perturbation-theory based solutions
\cite{tassev12,wang17,shi18}. The extension of such methods to biased
tracers \cite{wang18-recon} and including RSD \cite{zhu18-recon}, have
also been recently considered. Clearly, for the next generation of
experiment, reconstruction will be improved compared to the algorithms
used for BOSS, and we will have many methods to choose from.

\section{Conclusions}

This update on the lecture notes I provided 5 years ago at a previous
graduate school in Varenna \cite{percival13} clearly shows that the
best analysis method applied to measure clustering in galaxy surveys
continues to change, with better techniques being developed alongside
improvements in the experiments themselves. The techniques being used
in analyses now are very different and more robust than those used 5
years ago, and are resulting in more accurate measurements. Given the
excellent data becoming available in the next few years from DESI and
Euclid, it is clear that there is a strong driver for techniques to
continue to be developed. I look forward to writing the next set of
notes in 5 years time, if I'm invited back to lecture again at a
Varenna school.

\acknowledgments The author acknowledges useful conversations with
Davide Bianchi, Cullan Howlett \& Faizan Mohammad. 

\bibliographystyle{varenna}
\bibliography{percival-bib}

\end{document}